\documentclass{aastex62}

\newcommand{\Hmolion}{\textrm{H}_2^+}
\newcommand{\Hmol}{\textrm{H}_2}
\newcommand{\Hatom}{\textrm{H}}

\newcommand{\Heion}{\textrm{He}^+}
\newcommand{\Heionstate}{\textrm{He}^+(^2\textrm{S})}
\newcommand{\He}{\textrm{He}}
\newcommand{\Hestate}{\textrm{He}(1s^2\;^1\textrm{S})}
\newcommand{\Ar}{\textrm{Ar}}
\newcommand{\Arion}{\textrm{Ar}^+}
\newcommand{\Arstate}{\textrm{Ar}(^1\textrm{S})}
\newcommand{\Arionstate}{\textrm{Ar}^+(3p^5\;^2\textrm{P}^\mathrm{o})}
\newcommand{\HeArion}{\textrm{HeAr}^+}
\newcommand{\HeNeion}{\textrm{HeNe}^+}
\newcommand{\Arl}{{}^{36}\textrm{Ar}}

\newcommand{\ArHion}{\textrm{ArH}^+}
\newcommand{\ArHionl}{{}^{36}\textrm{ArH}^+}
\newcommand{\ArHionh}{{}^{38}\textrm{ArH}^+}
\newcommand{\Xstate}{\textrm{X}\,{}^2\Sigma^+}
\newcommand{\Astate}{\textrm{A}\,{}^2\Pi}
\newcommand{\Bstate}{\textrm{B}\,{}^2\Sigma^+}
\newcommand{\Ne}{\textrm{Ne}}
\newcommand{\rateunit}{\textrm{cm}^3/\textrm{s}}
\newcommand{\invcm}{\textrm{cm}^{-1}}

\received{\today}
\revised{\today}
\accepted{\today}
\submitjournal{ApJ}

\shorttitle{Radiative Charge Transfer Between $\Heion$ and $\Ar$}
\shortauthors{Babb \&  McLaughlin}

\begin{document}

\title{Radiative Charge Transfer Between the Helium Ion and Argon}

\correspondingauthor{James F. Babb}
\email{jbabb@cfa.harvard.edu,  bmclaughlin899@btinternet.com}

\author[0000-0002-3883-9501]{James F. Babb}
\affiliation{Harvard-Smithsonian Center for Astrophysics,
       MS 14, 60 Garden St., Cambridge, MA 02138-1516}

\author[0000-0002-5917-0763]{Brendan M. McLaughlin}
\affiliation{Centre for Theoretical Atomic Molecular and Optical Physics,
       Queens University Belfast,  Belfast BT7 1NN UK}



\begin{abstract}
The rate coefficient for radiative charge transfer between the helium ion
and an argon atom is calculated.
The rate coefficient is about $10^{-14}\;\rateunit$ at 300~K in agreement
with earlier experimental data.
\end{abstract}



\section{Introduction}

The molecular ion $\ArHionl$ (argonium)  
was identified through its rotational transitions in the sub-millimeter wavelengths
toward the Crab Nebula using SPIRE onboard the \textit{Herschel Space Observatory}~\citep{BarSwiOwe13},
with earlier and later observations 
revealing $\ArHionl$ and $\ArHionh$  in
the  diffuse interstellar medium using \textit{Herschel}/HIFI~\citep{SchNeuMul14}.
With the ALMA Observatory, the red-shifted spectra of $\ArHion$ was detected
in an external galaxy~\citep{MulMulSch15}.
Chemical modeling indicates that the presence of $\ArHion$ may serve
as a tracer of atomic gas
in the very diffuse interstellar medium and as an indicator of 
cosmic ray ionization rates~\citep{MulMulSch15,NeuWol16}.

$\Ar$ created in the supernova is predominately $\Arl$
and it was the isotope differences of $\ArHionl$ absorption spectra
compared to $\ArHionh$ absorption spectra 
that provided the crucial clues leading to the 
initial identification in the Crab nebula~\citep{BarSwiOwe13}.
Recently, \citet{PriBarVit17} modeled $\ArHion$ formation
in the filaments of the Crab nebula, considering the specific high-energy
synchrotron radiation found there, in addition to the cosmic ray flux considered earlier
for applications to the ISM.
A detailed exploration of model parameters led \citet{PriBarVit17} to the conclusion
that in order to reproduce observed abundances the relevant cosmic ray ionization
rate is $10^7$ times larger than the standard interstellar value,
though sensitivity to other parameters, such
as to the rate for dissociative recombination of $\ArHion$, was noted.

In models of the Crab Nebula $\Arion$ ions are produced through cosmic 
ray, UV and X-ray ionizations and by
the reaction of $\Hmolion$ with $\Ar$~\citep{Jen13,RouAleLeB14}.
The  $\ArHion$ is formed by the
ion-atom interchange reaction~\citep{BarSwiOwe13}
\begin{equation}
\Arion + \Hmol \rightarrow \ArHion + \Hatom 
\end{equation}
and destroyed by reaction with $\Hmol$~\citep{BarSwiOwe13,RouAleLeB14}.
Sources of $\Arion$ that enter the chemical models
were detailed by \citet{SchNeuMul14} and \citet{PriBarVit17}
and include photoionization of $\Ar$ and photodissociation of $\ArHion$. 
As \ion{He}{2} is present in the Crab nebula~\citep{SanHesSch98,Hes08}
it is fitting to investigate whether an additional source of $\Arion$ might
arise from charge transfer in Ar and $\Heion$ collisions.
Previous studies~\citep{SmiFleYou70,Isl74,AlbWir77} indicate that the direct charge transfer process
\begin{equation}
\label{CT}
\Arstate + \Heionstate \rightarrow \Arionstate + \Hestate 
\end{equation}
occurs through mechanisms involving couplings to excited molecular
states that are inaccessible
at thermal energies and
we conclude that 
charge transfer at thermal energies will proceed by the radiative charge transfer (RCT) 
mechanism
\begin{equation}
\Ar + \Heion \rightarrow \Arion + \He +h\nu,
\label{RCT-rxn}
\end{equation}
where $h\nu$ is the photon energy.
Experimentally, the  rate coefficient for the loss of helium ions in argon gas
was found by
\citet{JohLeuBio73} to be no more than $ 10^{-13}\;\rateunit$ at 295~K,
while \citet{JonLisTwi79} 
found that the rate coefficient  is greater than $2\times 10^{-14}\;\rateunit$ at 300~K;
both consistent with the early estimate of 
less than $ 10^{-13}\;\rateunit$  at 300~K 
\citep{FehSchGol66}.
Thus, (\ref{RCT-rxn}) is likely to be relatively slow,
but 
we are unaware of any detailed calculations for the cross sections
and rate coefficients of (\ref{RCT-rxn}).

\section{Molecular Calculations}
\begin{figure}[ht]
\epsscale{1.3}
\plotone{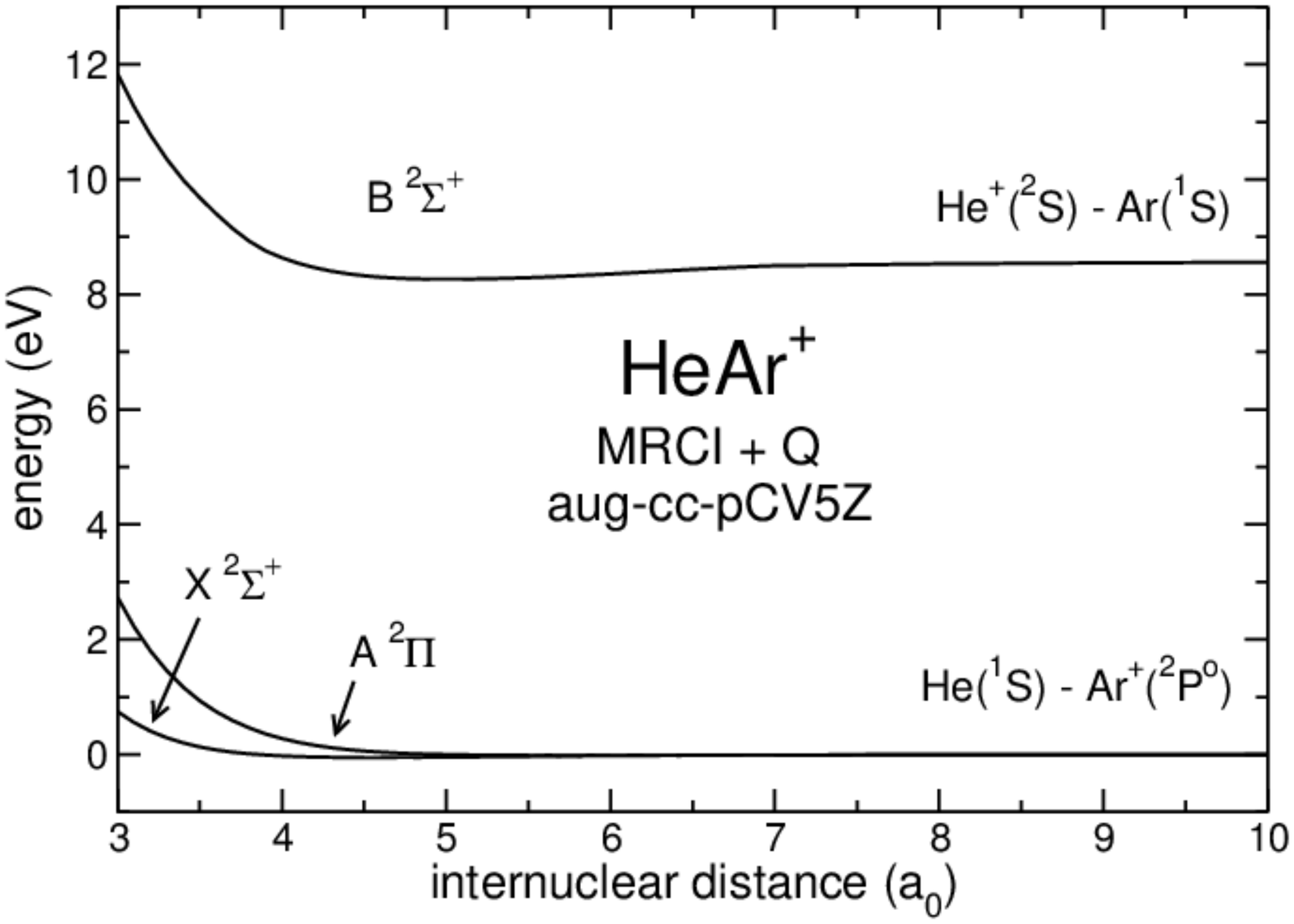}
\caption{Potential energy curves for the X$^2\Sigma^+$, A$^2\Pi$
         and B$^2\Sigma^+$ states of the HeAr$^+$ molecular cation as 
         a function of internuclear distance $R$ (in $a_0$). The results shown 
         were obtained in the MRCI + Q approximation using the 
         \textsc{molpro} quantum chemistry package of codes 
         with aug-cc-pCV5Z basis sets for He and Ar.}
\label{fig:pots}
\end{figure}

The molecular potential energy curve (PEC) of $\HeArion$ corresponding
to  the initial $\Arstate$ atom and $\Heionstate$ ion 
is the $\Bstate$ state, which lies 8.225 eV~\citep{StaPey86} above
the  $\Xstate$  and the $\Astate$ electronic states which 
correlate to  the final products 
$\Arionstate$ and $\Hestate$, see Fig.~\ref{fig:pots}.
Calculations of the $\Xstate$ and $\Astate$ PECs
were given by \citep{OlsLiu78,GemPey90,Sta90},
the last two including fine-structure for the $\Astate$ state,
and the $\Xstate$, $\Astate$, and $\Bstate$ states 
were calculated by \citet{LiaBalCha87}, using a relativistic CI method,
and by \citet{GemdeVPey90} using the MRD-CI method.
A summary  of earlier calculations is given by \citet{VieVigMas91}.
Semi-empirical functions fitting the PECs are available~\citep{Sis86,VieVigMas91}.
The $\Bstate$ state has a well depth of about 0.17~eV ($1375\;\invcm$),
while the $\Xstate$ and $\Astate$ states have very shallow potentials wells of
the order tens of meV~\citep{DabHerYos81,LiaBalCha87,VieVigMas91,CarLeaMar95}.
For the $\Xstate$ state the well depth is $286\;\invcm$, while
for the two states comprising the $\Astate$ (including fine structure) state
the depths are $\leq 182\;\invcm$.

\begin{figure}[ht]
\epsscale{1.30}
\plotone{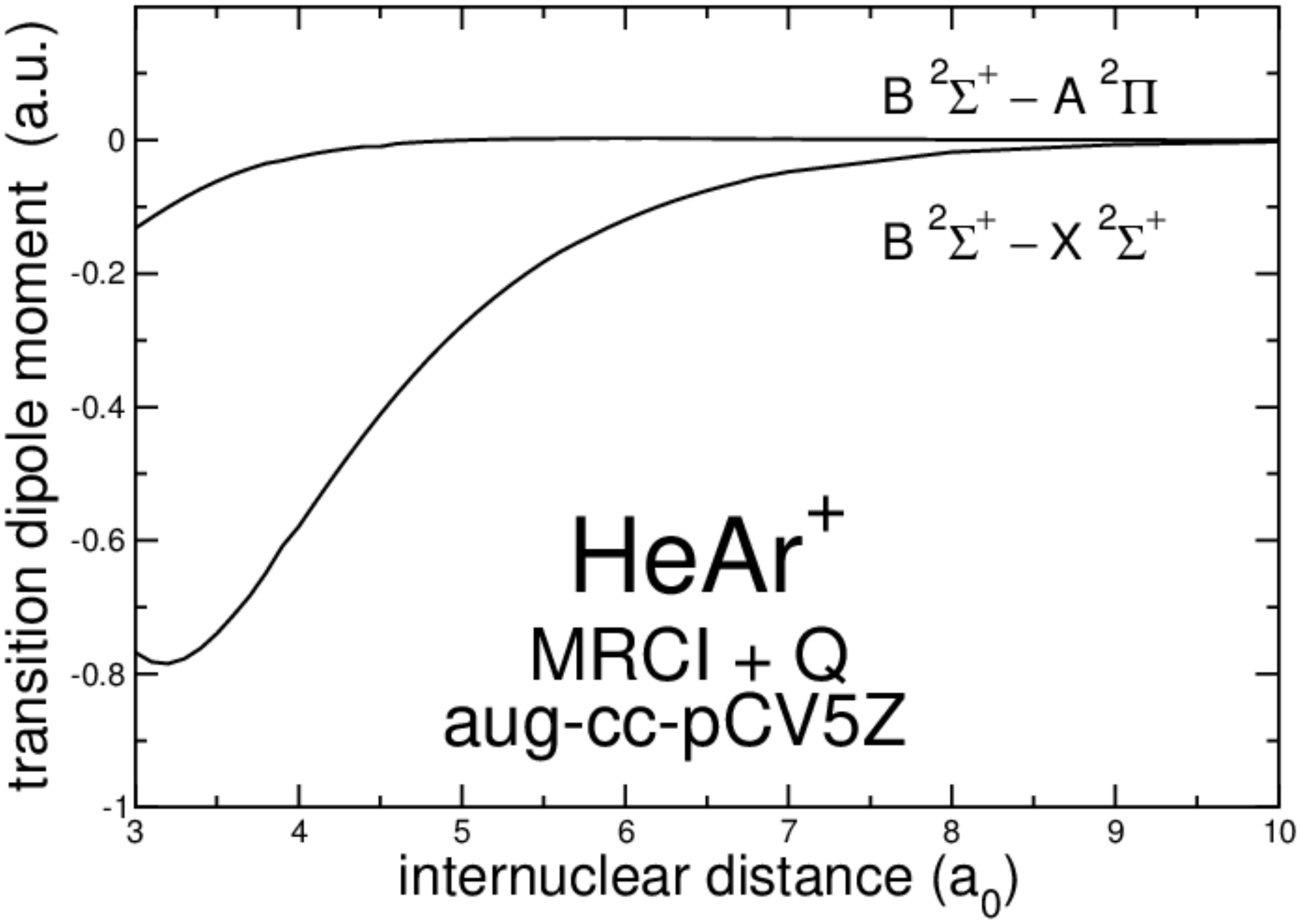}
\caption{Transition dipole moments for the B$^2\Sigma^+$--A$^2\Pi$
         and B$^2\Sigma^+$--X$^2\Sigma^+$ transitions in atomic units (a.u.) for the HeAr$^+$ molecular cation as 
         a function of internuclear distance $R$ (in $a_0$). 
         }
\label{fig:tdm}
\end{figure}

Depending on the relative magnitude of the transition dipole moments
for the $\Bstate$--$\Xstate$ and $\Bstate$--$\Astate$ transitions,
qualitatively, we might expect a relatively large
radiative charge transfer rate coefficient
because the $\Bstate$ state is relatively shallow,
the Franck-Condon overlap factors
governing matrix elements for
transitions between the $\Bstate$ state and
between the $\Astate$ state and the $\Xstate$ state
are favorable, 
and the net energy between the initial and final states is $\sim 8$~eV,
which enters the transition probability as a cubic factor.
To our knowledge the transition dipole moments coupling the
$\Bstate$ and $\Xstate$ states and
the  $\Bstate$ and $\Astate$ states are not available in the literature, 
so we calculated them using the latest version of the 
quantum chemistry package  \textsc{molpro}, running on parallel computer architectures.
The present results, described in more detail below, are shown in Fig.~\ref{fig:tdm}.
Inspection of the calculated transition dipole
moments indicates that the leading channel
for radiative charge transfer will be 
the $\Bstate$--$\Xstate$ transition in comparison
to the  $\Bstate$--$\Astate$  transition,
similarly to the case of radiative charge transfer between
$\Heion$ and $\Ne$~\citep{LiuQuXia10}.

Potential energy curves (PECs) and transition dipole moments (TDMs) as a 
function of internuclear distance $R$ were calculated for
a range of values of internuclear distance $R$ between 3 and $60\;a_0$.
We used a 
state-averaged-multi-configuration-self-consistent-field (SA-MCSCF) approach, 
followed by multi-reference
configuration interaction (MRCI) calculations together with the Davidson correction 
(MRCI+Q)~\citep{Helgaker00},
in a similar manner to our recent molecular structure work on the HeC$^+$ and
the CH$^+$ molecular complexes \citep{Babb17a,Babb17b}.
The SA-MCSCF method is used as the reference wave function for the MRCI calculations. 
All the molecular data were obtained with 
\textsc{molpro} 2015.1 \citep{MOLPRO_brief}
in the C$_{2v}$ Abelian symmetry point group
using  augmented - correlation - consistent 
polarized core-valence quintuplet basis sets (aug-cc-pCV5Z) for each atom/ion, 
in the MRCI+Q calculations \citep{Helgaker00}. 
The use of large basis sets in molecular electronic structure calculations 
are well known to recover $\approx$ 
98\% of the electron correlation effects. 

In detail, for the $\HeArion$ molecular cation, thirteen 
molecular orbitals (MOs) are put into the
active space, including seven $a_1$, three $b_1$ and three $b_2$ symmetry
MOs. The rest of the electrons in the $\HeArion$ system are
put into the closed-shell orbitals. The MOs for the MRCI procedure are
obtained from the SA-MCSCF method, where the averaging
process was carried out on the lowest four $^2\Sigma^+$ ($^2A_1$), 
four $^2\Pi$ ($^2B_1$), and four $^2\Delta$ ($^2A_2$) molecular states of this molecule.
We then use these thirteen MOs (7$a_1$, 3$b_1$, 3$b_2$, 0$a_2$), i.e. (7,3,3,0), to perform
all the PEC calculations of these electronic states in the MRCI + Q approximation as 
a function of bond length.
The potential energies (PECs) and transition dipole moments (TDMs), respectively, are shown in
Figs.~\ref{fig:pots} and \ref{fig:tdm}, for the restricted range $3<R<10\;a_0$.

The long-range form of the $\Bstate$ state
potential energy, correlating to $\Heion ({}^2 S_0)$-$\Ar ({}^1S_0)$, is (in atomic units)
\begin{equation}
V_B (R) \sim  -\textstyle{\frac{1}{2}} \alpha (\Ar) R^{-4} -\frac{1}{2} \alpha_q (\Ar)R^{-6}  - C_6 (\Heion\cdot\Ar)R^{-6},
\end{equation}
where the electric
dipole and quadrupole polarizabilities of Ar
and the dispersion (van der Waals) constant of $\Heion\cdot\Ar$
are, respectively, $\alpha (\Ar) = 11.08$~\citep{KumTha10}, $\alpha_q (\Ar) = 52.8$~\citep{JiaMitChe15}, and $C_6 = 2.36$.
We calculated $C_6$ using the 
oscillator strength distributions~\citep{Bab94} of $\Heion$~\citep{JohEpsMea67}
and of $\Ar$~\citep{KumTha10}.
[The present value of $\frac{1}{2} \alpha_q (\Ar)  + C_6 (\Heion\cdot\Ar)= 28.8$ and is in
good agreement with value 27.6 given by~\citet{Sis86}, who estimated $C_6$ using
the Slater-Kirkwood approximation.]

The long-range forms of the $\Xstate$ and $\Astate$ states, correlating
to $\Hestate$-$\Arionstate$, are (in atomic units)
\begin{equation}
V_{A;X} (R) \sim -\textstyle{\frac{1}{2}}\alpha(\He)  R^{-4} -8.64 R^{-6} ,
\end{equation}
where the electric dipole
polarizability of He is $\alpha(\He)= 1.383$~\citep{YanBabDal96} and the term  ${\cal O}(R^{-6})$ is 
an estimate  of the contribution of the 
electric quadrupole polarizability of helium and the 
dispersion constant 
for $\He\cdot\Arion$  \citep{Sis86,CarLeaMar95}.
The transition dipole moments are fitted to the form ${R}^{-12}$ for $R>16\;a_0$.

The calculated $\Bstate$ potential gives a well of depth $D_e=0.03132$~eV $(2526\;\invcm)$ at $R=5\;a_0$, 
to be compared to the experimentally determined well depth of  $1375\;\invcm$
at the same value of $R$~\citep{DabHerYos81}.
Part of this discrepancy may be due to our neglect of fine structure (spin-orbit coupling),
as shown for the $\Xstate$ and $\Astate$ states by \citet{Sta90}.

\section{Cross sections and rate coefficients}

The radiative loss cross sections were calculated using the optical potential approach,
which accounts for radiative loss in collisions of $\Heion$  with $\Ar$.
The theory is described in~\citet{Babb17a},
see also, for example, \citet{ZygDal88,LiuQuWan09} and references therein.
In carrying out the calculations of (\ref{RCT-rxn}), the probability of approach in the $\Bstate$ state
is unity. The reduced mass of the colliding system corresponds to ${}^4\He$ and $\Arl$.

\begin{figure}[ht]
\epsscale{1.}
\plotone{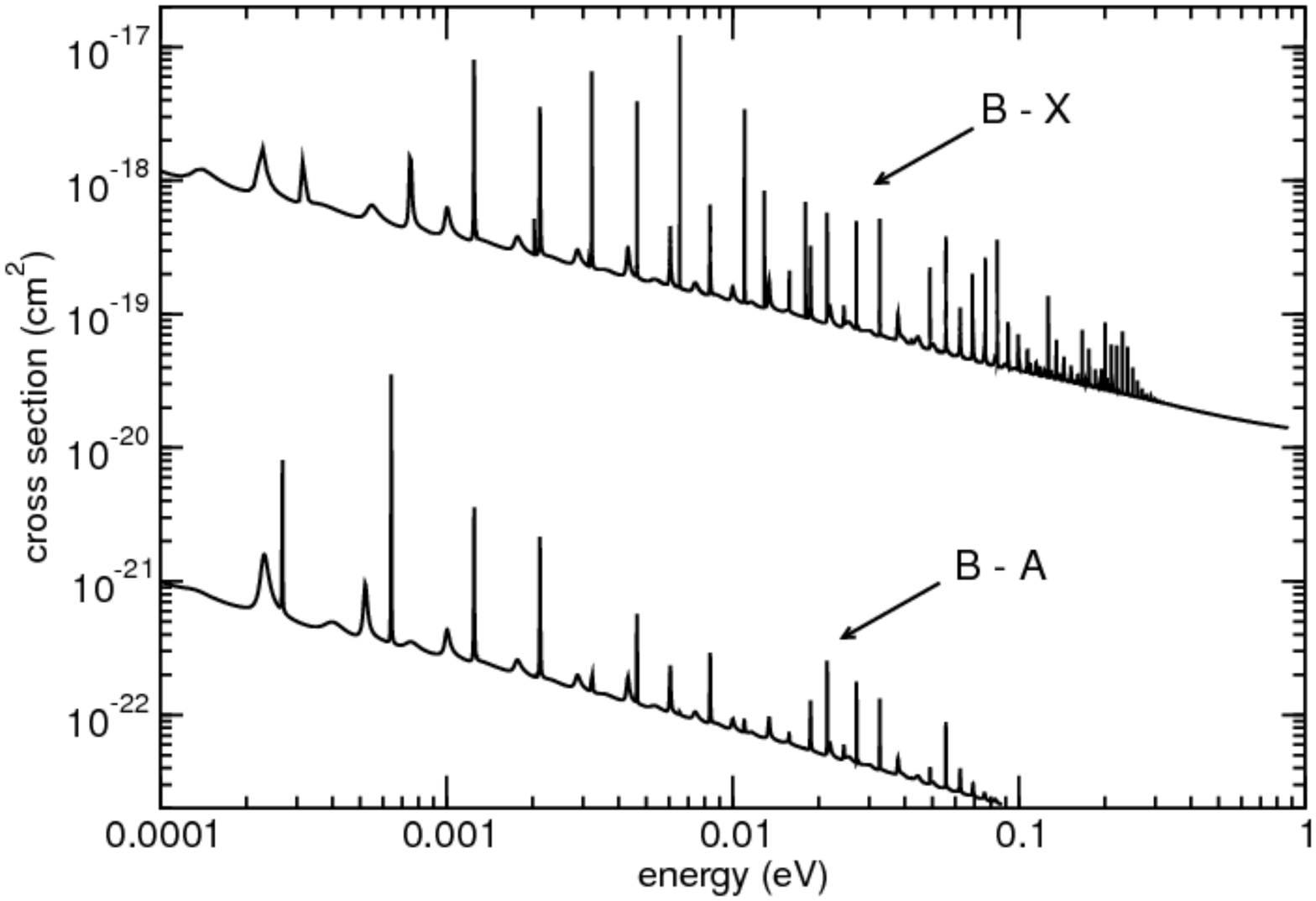}
\caption{Cross sections in $\textrm{cm}^2$ for radiative loss in
         collisions of $\Heion$ with $\Ar$. Upper curve
         for $\Bstate$ to $\Xstate$ transitions,
         lower curve for $\Bstate$ to $\Astate$ transitions.
         }
\label{fig:cross}
\end{figure}

The calculated cross sections for radiative loss (radiative charge transfer)
are shown in Fig.~\ref{fig:cross}.
As expected, the $\Bstate$ to $\Xstate$ transition is significantly
larger than the $\Bstate$ to $\Astate$ transition.
The radiative loss cross sections include the possibility of radiative association;
however, because the well depth of the $\Xstate$ state  is only 262~$\invcm$
and the well depth of the $\Astate$ state is even less~\citep{DabHerYos81,CarLeaMar95}, 
and because the transition dipole moments are diminished
at the equilibrium distances characteristic of these states (about $5\;a_0$),
the  relative contributions
of radiative association to radiative loss
are assumed to be very small.
Cross sections for loss through radiative association from the initial $\Bstate$ state 
to bound levels of the $\Bstate$ state are also insignificant compared
to the cross sections given in Fig.~\ref{fig:cross} because
the transition probabilities depend on the de-excitation energies to the third power~\citep{ZygLucHud14}.
Earlier detailed calculations on the $\textrm{LiHe}^+$ system 
are illustrative~\citep{DalKirSta96,StaZyg96},
in 
particular, Figure 2 of \citet{AugSpiKra12}.

We therefore take the radiative loss cross section as a good
approximation for the radiative charge transfer cross section.
This is in contrast to the $\HeNeion$ system,
where the well depth of the ground state is $D_e=6216\pm 300\;\invcm$
and the transition dipole moments ($\Bstate$--$\Xstate$ and $\Bstate$--$\Astate$) are comparable,
with  roughly equal contributions to radiative loss from 
radiative charge transfer and from radiative association~\citep{CooKirDal84,LiuQuWan09}.
We note that cross sections for the various transitions that 
originate from the excited doublet electronic state
of HeAr$^{+}$ have Langevin $1/v$ (or $E^{-1/2}$), where
$v$ is the relative velocity (and $E$ is the relative kinetic
energy), background dependences at low energies, 
with overlying resonance features. The rate coefficients
for process (\ref{RCT-rxn}) were calculated by averaging the cross sections
for the $\Bstate$--$\Xstate$ transition over a Maxwellian velocity distribution.
The results are given in Table~\ref{tab:rates}.

To check the sensitivity of the cross sections to the description of
the $\Bstate$ state, we  repeated the calculations with the empirically
determined $\Bstate$ potential of \citet{Sis86}, which reproduces the experimental
well depth.
The main effect of using the empirical $\Bstate$ state potential
was to change the positions of the resonances in the cross section at low energies, but
this did not appreciably affect
the values of the rate coefficients at thermal energies.
We verified that 
the $\Bstate$ to $\Xstate$ cross sections
shown in Fig.~\ref{fig:cross} satisfied the alternative upper
bound for radiative charge transfer cross sections given in Eq.~(A.6)
of~\citep{ZygLucHud14}. The upper bound
was consistently 
larger; by about 4\% at energy $9\times 10^{-4}$~eV
increasing to about 9\% at energy 0.09~eV.

\startlongtable
\begin{deluxetable}{CCC}
\tablecaption{Rate coefficients ($\rateunit$) for
radiative charge transfer in collisions
of $\Heion$ with $\Ar$. Numbers in parentheses
represent powers of 10 by which the preceding
number should be multiplied.
\label{tab:rates}}
\tablehead{
\colhead{Temperature} & \colhead{Rate coefficient} &  \\
\colhead{K} & \colhead{$\rateunit$} &  
}
\startdata
        77   &  1.05($-$14) &\\
       100   &  1.05($-$14) &\\
       200   &  1.01($-$14) &\\
       300   &  9.86($-$15) &\\
       400   &  9.69($-$15) &\\
       500   &  9.56($-$15) &\\
       600   &  9.48($-$15) &\\
       700   &  9.39($-$15) &\\
       800   &  9.32($-$15) &\\
       900   &  9.28($-$15) &\\
      1000   &  9.30($-$15) &\\
      2000   &  8.89($-$15) &\\
      3000   &  8.33($-$15) &\\
\enddata
\end{deluxetable}

\section{Discussion}
The calculated cross sections and rate coefficients
are somewhat larger than those
for radiative charge transfer between
$\Heion$ and $\Ne$.
For example, 
for radiative charge transfer of $\Heion$ and
$\Ne$ at 300~K
experiment gives $1.0(3) \times 10^{-15}\;\rateunit$~\citep{Joh83}
to be compared with the theoretical value 
of about $5\times 10^{-16}\;\rateunit$ \citep{CooKirDal84,LiuQuWan09},
while for radiative charge transfer
of $\Heion$ and $\Ar$, we find a
rate coefficient of 
$9.86\times 10^{-15}\;\rateunit$.
Similarly, at 77~K \citet{Joh83} measured
a rate coefficient $\sim 2\times 10^{-15}\;\rateunit$ for
radiative charge transfer of $\Heion$ and
$\Ne$ and \citet{LiuQuWan09} calculate $3\times 10^{-16}\;\rateunit$,
while at the same temperature for 
$\Heion$ and $\Ar$ we find a rate coefficient
of $1.05\times 10^{-14}\;\rateunit$.
The larger calculated values for 
the rate coefficients in the $\Ar$ system, compared
to the calculated values for the $\Ne$ system, arise from 
two opposing effects.
The dissociation limit of
the $\Bstate$ state 
is about 8.225~eV above the $\Xstate$ state
for $\HeArion$ and
compared to 
about 3.125~eV for $\HeNeion$, while
the transition dipole moment
of the dominant 
$\Bstate$ to $\Xstate$ transition in
$\HeNeion$~\citep{CooKirDal84,LiuQuWan09}
is slightly larger than
that for $\HeArion$.

However, the rate coefficient for 
process (\ref{RCT-rxn})
is probably not large enough
to be a significant factor in the production
of $\ArHion$ in the diffuse interstellar medium.
The corresponding radiative charge transfer reaction has been added to the chemical 
network relevant to the $\ArHion$ molecular ion, as described in~\citet{NeuWol16} 
in the PDRLight version of the Meudon PDR code \citep{LeBLePPin12}.\footnote{Available at http://ism.obspm.fr}
The model was run both for diffuse galactic cloud conditions 
(standard interstellar radiation field, 
$n_\mathrm{H} =50\;\textrm{cm}^{-3}$, $T=50$~K, low visual extinction, 
cosmic ionization rate $\zeta = 10^{-16}\;\textrm{s}^{-1}$) 
and for conditions corresponding to the Crab Nebula environment 
as reported in \citep{PriBarVit17} ($n_\mathrm{H}=2000\;\textrm{cm}^{-3}$,  $\zeta = 10^{–9}\;\textrm{s}^{-1}$).
Using the value $10^{-14}\;\rateunit$
for the rate coefficient
of the radiative charge transfer reaction (\ref{RCT-rxn}), no significant modifications 
of the chemical equilibrium of the $\ArHion$ molecular ion 
were found for either case.

\acknowledgments
We thank Dr. Evelyne Roueff for suggesting this investigation and for model calculations
using PDRLight.
B.M.McL. acknowledges support 
from the Institute for Theoretical Atomic,
Molecular, and Optical Physics (ITAMP) visitors program,
from Queen's University Belfast
for the award of a visiting research fellowship (VRF)
and from a Smithsonian Institution Scholarly Studies award.
ITAMP is supported in part by a grant from the NSF to the Smithsonian
Astrophysical Observatory and Harvard University. 
The computational work was performed at the National Energy Research Scientific
Computing Center in Berkeley, CA, USA and at The High Performance 
Computing Center Stuttgart (HLRS) 
of the University of Stuttgart, Stuttgart, Germany. 
Grants of computing time at NERSC and HLRS are gratefully acknowledged. 

\software{molpro 2015.1, \citep{MOLPRO_brief}; PDRLight, \citep{LeBLePPin12}  }

\end{document}